\documentclass[a4paper,12pt,showpacs,preprintnumbers]{elsarticle}
\usepackage{amsmath,amssymb}
\usepackage{slashed}
\usepackage{mathptmx}
\usepackage{url}
\allowdisplaybreaks 
\newcommand{\beq}{\begin{equation}}
\newcommand{\eeq}{\end{equation}}

\newcommand{\be}{\begin{eqnarray}}
\newcommand{\ee}{\end{eqnarray}}
\begin{document}
\title{On transverse spin sum rules}
\author{A. Harindranath}
\ead{a.harindranath@saha.ac.in}
\address{Theory Division, Saha Institute of Nuclear Physics,
1/AF Bidhan Nagar, Kolkata 700064, India}
\author{Rajen Kundu}
\address{Department of Physics, RKMVC College,
Rahara, Kolkata,  700118, West Bengal, India }
\ead{rajenkundu@yahoo.com}
\author{Asmita Mukherjee}
\address{Department of Physics, Indian Institute of Technology Bombay,
Powai, Mumbai 400076, India}
\ead{asmita@phy.iitb.ac.in}
\date{29 July 2013}

\begin{abstract}
In this work we provide explicit calculations that support the
conclusions stated in Ref. \cite{hkmr} regarding recent literature on
transverse polarization. We also compare and contrast two methods of
deriving spin sum rules. 
\end{abstract}
\maketitle
{\bf Introduction}
\vskip .2in
At present, understanding the helicity and transverse spin structure of the
proton in the context of Deep Inelastic Scattering (DIS) is of great
interest. Intense experimental and theoretical research
activities have been going on in this field for more than a decade.
It is well-known that since DIS is a light-cone dominated process, the most
appropriate theoretical tool to study it is provided by Light Front 
Quantization (for a review, see Ref. \cite{bpp}). In order to understand the
spin structure of proton which is a composite object and
investigate any sum rule associated with it, one should start from the
intrinsic spin operators ${\cal J}^i, i=1,2,3$ which can be constructed
from the Pauli-Lubanski operator. It is well-known that ${\cal J}^i$'s are
frame independent (see for example, Refs. \cite{bh,soper,ls}) whereas 
the usual rotation operators (which form part of 
the Poincare generators) are frame dependent. Any angular momentum sum rule, 
solely based on rotation operators
that are part of Poincare generators, will have frame dependence. 
 Same is also true, in general, if one starts with Pauki-Lubansky operators 
as we discuss below. As
already stated, the
solution to this problem is to start from intrinsic spin operators ${\cal
J}^i$.

Helicity operator ${\cal J}^3$ (whose explicit construction and a perturbative
analysis in light front QCD is carried out in Ref. \cite{hk} in the 
total transverse momentum zero frame) is kinematical (interaction free).
On the other hand, it is well known that the transverse rotation operators 
and hence the transverse spin operators in light front theory are dynamical 
(interaction dependent). Construction and analysis of ${\cal J}^i,~ (i=1,2)$
in light front QCD is carried out in Ref. \cite{hmr1,hmr2}.

We have shown in Ref. \cite{hmr1,hmr2} that just like the helicity operator
${\cal J}^3$, the transverse spin ${\cal J}^i,~i=1,2$ of the
composite  state can be
separated in the gauge $A^+=0$ into orbital-like (explicit dependence on the
coordinate
$x^\perp$) contribution ${\cal J}^i_{I}$ and coordinate-independent  parts
${\cal J}^i_{II}$ and ${\cal J}^i_{III}$. What is the phenomenological
relevance of this separation? Most interestingly, the proton
matrix element of ${\cal J}^i_{II}$ is shown to be directly related to the
integral of the well-known transverse polarized structure function 
$g_T$ just as the proton matrix element of the coordinate-independent
quark intrinsic part of 
${\cal J}^3$ is related to the polarized structure function $g_1$. Based on
${\cal J}^i,~ i=1,2$ in light front QCD, in Ref. \cite{hmr2}, a transverse  
spin sum rule was proposed  and  verified for a dressed quark to 
${\cal O}(g^2)$ in perturbation theory.


Recently, the matrix element of the transverse component of the 
Pauli-Lubanski operator has been formally analyzed in Refs. \cite{jxyplb}
and \cite{jxy}
following the approach of Ref. \cite{jm}
and using  the parameterizations of
the off-forward matrix elements of the energy-momentum tensor.
Refs. \cite{jxyplb} and \cite{jxy}  appear to be partly inspired by
Ref. \cite{bur} in which a relation between the expectation
value of equal time transverse  rotation generator $J_q^i$ and the form
factors 
$A_q(0)$ and $B_q(0)$ is
obtained using
 delocalized wave packet states that are transversely polarized in the rest
frame of the
nucleon.    

We have pointed out in Ref. \cite{hkmr} that many of the statements
 in Refs. \cite{jxyplb} and \cite{jxy} appear unsupported by explicit
calculations. In this work we present explicit calculations supporting our
statements in Ref. \cite{hkmr}. We also compare and contrast our method
\cite{hk,hmr1,hmr2} with the method used in Refs. \cite{jxyplb} and
\cite{jxy} to derive sum rules.


\vskip .2in
{\bf General outline of the calculation} 
\vskip .2in
The starting point in Ref. \cite{jxyplb} and \cite{jxy} is the Pauli-Lubanski operator which is defined in terms of energy 
momentum tensor in a very standard way as follows.
\be
W^\mu & = & -\frac{1}{2}\epsilon^{\mu\nu\alpha\beta}J_{\nu\alpha}P_{\beta}
\nonumber\\
M^{\mu \nu} & = & \frac{1}{2} \int dx^- d^2x^\perp ~ 
\Big [ x^\mu T^{+ \nu} - x^\nu T^{+ \mu} \Big ]~.
\ee

 It is to be noted that in Refs. \cite{jxyplb} and \cite{jxy} 
the authors have calculated the matrix element of  $W^i$ in the frame 
 $P^\perp=0$ (which is clear from Eq. (9) of \cite{jxyplb}), 
although the results have been claimed to be frame independent.  
On the other hand, in Ref. \cite{hk,hmr1,hmr2} where calculations are done completely within the framework of light-front QCD using Light-front gauge, intrinsic spin operators are used. Note that intrinsic helicity operator ${\cal J}^3$ and intrinsic transverse spin ${\cal J}^i, i=1,2$ among themselves obey the angular momentum algebra (see for example, Refs. \cite{bh,soper,ls}).  For a massive particle like nucleon, intrinsic spin operators and Pauli-Lubanski operators are related through the following relations.
\be
M {\cal J}^i  &=& W^i - P^i {\cal J}^3 
       = \epsilon^{ij}(\frac{1}{2} F^j P^+ + K^3 P^j- \frac{1}{2} E^j P^-)
         - P^i {\cal J}^3 ~,
\nonumber\\
{\cal J}^3 &=& \frac {W^+}{P^+} = J^3 + \frac{1}{P^+}(E^1 P^2 -  E^2 P^1)
\ee
In the above expressions, $F^i=M^{-i}$ are the light-front transverse 
rotation operators and are interaction dependent or dynamical; 
while $E^i=M^{+i}$ are light-front transverse boost operators and are 
interaction independent or kinematical. Longitudinal boost $K^3=M^{+-}$ and 
helicity $J^3=M^{12}$ are also kinematical. Note that the light front 
transverse rotation and boost operators were
mis-identified in Refs. \cite{jxyplb} and \cite{jxy}. This was already
pointed out in Ref. \cite{ll}.  
Moreover, the authors of Refs. \cite{jxyplb} and \cite{jxy} did not consider 
longitudinal boost operator $K^3=M^{+-}$ for working explicitly 
in $P^\perp=0$ frame and only for such a choice of frame, both the starting 
points appear to be the same.  In the following, we kept this term to show 
an 
example in the course of our explicit calculations that, in general, for a 
frame with non-zero $P^\perp$ both are not the same. 
Following Refs. \cite{jxyplb} and \cite{jxy} we also assume that the various
Poincare generators can be separated to quark and gluon parts.

Next, to compare with the results of Refs. \cite{jxy} and \cite{jxyplb},  
we need 
to calculate the transverse component of the Pauli-Lubanski operator
corresponding to species $i$ formally defined as
\be 
W_i^1 = \frac{1}{2} F_i^2 P^+ + K_i^3 P^2- \frac{1}{2} E_i^2 P^-
\label{w1}\ee
 and its matrix element in a
transversely polarized state
\be
\frac{\langle P S^{(1)} \mid W_i^1 \mid P S^{(1)} \rangle}
{(2
\pi)^3 2P^+ \delta^3(0)}~ \label{wspin}
\ee
where $i$ denotes either the quark or gluon part. 
Note that, in the rest of the paper, we always deal with only one component, 
namely, $W_i^1$, while calculation with $W_i^2$ is trivially same and 
unnecessary for our purpose. 

The transverse rotation operator is
\be
F_i^2 =  \frac{1}{2} M_i^{-2}  
  & = & \frac{1}{4} \int dx^- d^2x^\perp ~ \Big [
x^- ~ T_i^{+2} - x^2 ~T_i^{+-} \Big ]~.
\ee
We note that, 
\be 
K_i^3 =  \frac{1}{2} M_i^{+-}  
  & = & \frac{1}{4} \int dx^- d^2x^\perp ~ \Big [
x^+ ~ T_i^{+-} - x^- ~T_i^{++} \Big ]~\nonumber\\
& = & \frac{1}{2} x^+ P^- + {\tilde K}_i^3~,  \nonumber\\
E_i^2 =  M_i^{+2} 
  & = & \frac{1}{2} \int dx^- d^2x^\perp ~ \Big [
x^+ ~ T_i^{+2} - x^2 ~T_i^{++}     \Big ]~ \nonumber \\
& = & x^+ P^2 +{\tilde E}_i^2 ~.
\ee
In writing the last equalities in both the above expressions, 
we note that light-front time $x^+$ can be taken out of the integral in the 
first terms  and simplified. Putting them back in Eq. (\ref{w1}), 
we see that only the second terms in these expressions contribute to $W_i^1$.
Thus we find that 
\be
W_i^1 = \frac{1}{2} F_i^2 P^+ + {\tilde K}^3 P^2- \frac{1}{2} {\tilde E}_i^2
P^-~
\ee
with no explicit $x^+$ dependence.
Lastly, the light-front helicity operator  is given by
\be
J_i^3 =   M_i^{12}  
  & = & \frac{1}{2} \int dx^- d^2x^\perp ~ \Big [
x^1 ~ T_i^{+2} - x^2 ~T_i^{+1} \Big ]~.
\ee 

According to the procedure prescribed in Ref. \cite{jm} and 
followed in Refs. \cite{jxyplb,jxy}, rest of the calculation relies on defining 
the Fourier transform of the off-forward matrix elements of relevant 
component of energy momentum tensor and then consider the forward limit. 
Since $W_i^1$ is independent of $x^+$  explicitly, we consider 
three dimensional Fourier Transform of the off-forward
matrix element. In general, we define
\be
\langle P' S^{(1)}\mid   {\cal {\hat O}^\alpha} (k_-, k_i) 
\mid P S^{(1)} \rangle =
\frac{1}{2} \int dx^- d^2x^\perp ~ e^{i(k_- x^- + k_i x^i)}~x^\alpha 
~\langle P' S^{(1)}\mid {\cal O} (x)
\mid P S^{(1)} \rangle
\ee
where $ \alpha = -,1,2. $  
Using translational invariance, we find 
\be
\langle P' S^{(1)}\mid   {\cal {\hat O}^\alpha} (k)
\mid P S^{(1)} \rangle &=& - i (2 \pi)^3~\frac{\partial}{\partial k_\alpha}
\Big [ \delta^3(k+P'-P) ~\langle P' S^{(1)}\mid {\cal O} (0)
\mid P S^{(1)} \rangle \Big ] \nonumber \\
&=&- i (2 \pi)^3~\delta^3(k+P'-P)~ \frac{\partial}{\partial k_\alpha}
  ~\langle P' S^{(1)}\mid {\cal O} (0)
\mid P S^{(1)} \rangle
\ee
ignoring the term containing the derivative on the delta function \cite{jm}.

Thus, with $\Delta=P'-P$, we find
\be
\langle P S^{(1)}\mid  { F_i}^{2}  
\mid P S^{(1)} \rangle &=& i (2 \pi)^3 \delta^3(0) ~
\Big [ \frac{\partial}{\partial \Delta_-}  \langle P' S^{(1)} \mid T_i^{+2} (0)
\mid P S^{(1)} \rangle ~ \nonumber \\
&& - ~
\frac{\partial}{\partial \Delta_2} \langle P' S^{(1)} \mid T_i^{+-} (0) 
\mid P S^{(1)} \rangle 
\Big]_{\Delta=0}~,
\ee
\be
\langle P S^{(1)}\mid  {{{\tilde K}}_i}^{3} 
\mid P S^{(1)} \rangle =- \frac{i}{2} (2 \pi)^3 \delta^3(0) ~
\Big [ 
\frac{\partial}{\partial \Delta_-} \langle P' S^{(1)} \mid T_i^{++} (0)
\mid P S^{(1)} \rangle  \Big]_{\Delta=0}~.
\ee  
and 
\be
\langle P S^{(1)}\mid  { {{\tilde E}}_i}^{2} 
\mid P S^{(1)} \rangle =- i (2 \pi)^3 \delta^3(0) ~
\Big [ 
\frac{\partial}{\partial \Delta_2} \langle P' S^{(1)} \mid T_i^{++} (0)
\mid P S^{(1)} \rangle  \Big]_{\Delta=0}~.
\ee  
\eject
{\bf Matrix elements of the Energy-Momentum tensor}
\vskip .1in
We start from the following parameterization as used in Refs. \cite{jxy,jxyplb},
\be
\langle P',S' \mid T_i^{\mu \nu}(0) \mid PS \rangle & = &
{\overline U}(P',S') \Bigg 
[ A_i(\Delta^2) \frac{1}{2} \Big( \gamma^\mu {\overline P}^\nu +
 \gamma^\nu {\overline P}^\mu \Big) \nonumber \\
&~&~~~ + B_i(\Delta^2) \frac{1}{2 M_N} \frac{1}{2} \Big(
{\overline P}^\mu i \sigma^{\nu \alpha} \Delta_\alpha +
{\overline P}^\nu i \sigma^{\mu \alpha} \Delta_\alpha \Big ) \nonumber \\
&~&~~~ + C_i(\Delta^2) \frac{1}{M_N} \Big(\Delta^\mu \Delta^\nu - 
g^{\mu \nu} \Delta^2  \Big) + {\overline C}_i (\Delta^2) M_N g^{\mu \nu}
\Bigg] U(P,S). \nonumber \\
\label{eq1} \ee
Here ${\overline P} = \frac{1}{2}(P+P')$.

Note that, in the above parameterization, effects of QCD-interactions are 
buried in 
the form factors while the associated Lorentz structures 
are given in terms of asymptotic spin-half nucleonic states.
We can either calculate the matrix elements in Eq. (\ref{eq1}) directly
(which we denote by method I) or 
use the Gordon identity
\be
{\overline U}(P',S') \frac{i}{2 M_N} \sigma^{\mu \nu} \Delta_\nu U(P,S) =
{\overline U}(P',S') \gamma^\mu U(P,S) - {\overline U}(P',S')
\frac{(P+P')^\mu}{2M_N} U(P,S)~ \label{gc}
\ee
to eliminate either the $``$$\sigma$" terms (method II) or 
the $``$$\gamma$" terms (method III which is used in Ref. \cite{jxyplb}) 
from Eq. (\ref{eq1}). 
 
Eliminating the $``$$\sigma$" terms (method II) we have
\be
\langle P',S' \mid T_i^{\mu \nu}(0) \mid PS \rangle & = &
{\overline U}(P',S') \Bigg[  - B_i (\Delta^2) 
\frac{{\overline P}^\mu {\overline P}^\nu}{M_N}  \nonumber \\
&~&~+ (A_i(\Delta^2)+ B_i(\Delta^2)) \frac{1}{2} \Big (
\gamma^\mu {\overline P}^\nu  + \gamma^\nu {\overline P}^\mu\Big) \nonumber \\
&~&~~~ + C_i(\Delta^2) \frac{1}{M_N} \Big(\Delta^\mu \Delta^\nu - 
g^{\mu \nu} \Delta^2  \Big) + {\overline C}_i (\Delta^2) M_N g^{\mu \nu} 
\Bigg] U(P,S) \nonumber \\ 
\label{eq2}\ee
On the other hand, eliminating the $``$$\gamma$" terms (method III) we have
\be
\langle P',S' \mid T_i^{\mu \nu}(0) \mid PS \rangle & = &
{\overline U}(P',S') \Bigg 
[ A_i(\Delta^2)  \frac{{\overline P}^\mu {\overline P}^\nu}{M_N} \nonumber \\
&~&~~~ + (A_i(\Delta^2) +  B_i(\Delta^2)) \frac{1}{2 M_N} \frac{1}{2} \Big(
{\overline P}^\mu i \sigma^{\nu \alpha} \Delta_\alpha +
{\overline P}^\nu i \sigma^{\mu \alpha} \Delta_\alpha \Big ) \nonumber \\
&~&~~~ + C_i(\Delta^2) \frac{1}{M_N} \Big(\Delta^\mu \Delta^\nu - 
g^{\mu \nu} \Delta^2  \Big) + {\overline C}_i (\Delta^2) M_N g^{\mu \nu}
\Bigg] U(P,S). \nonumber \\
\label{eq3}\ee

All the three methods should yield the same results since Eq. (\ref{gc}) is
simply an identity which is valid for on-shell spin-half states. 

We present explicit calculation in Method II. Further details of the calculation are 
given in the appendix that are used to obtain the results given below.

In the following we keep only terms which are linear in $\Delta$, which are
relevant for the computation of matrix elements of transverse spin.
Then, the matrix elements of $T^{\mu \nu}(0)$ in the transversely polarized
state (to be specific, taken to be polarized along +ve $x$ direction) are
(in the frame ${\overline P}^\perp=0$)
\be
\langle P',S^{(1)} \mid T_i^{++}(0) \mid PS^{(1)} \rangle & = &
- B_i(\Delta^2) \frac{{\overline P}^+ {\overline P}^+}{M_N} 
\Big (  - i \Delta^{(2)}  \Big )~, \label{pp} \\
\langle P',S^{(1)} \mid T_i^{+1}(0) \mid PS^{(1)} \rangle & = & 0,
\label{p1}\\
\langle P',S^{(1)} \mid T_i^{+2}(0) \mid PS^{(1)} \rangle & = &
\frac{1}{2} \Big (A_i(\Delta^2) + B_i(\Delta^2)\Big )~(-i) ~M_N~ \Delta^+
~, \label{p2}\\ 
\langle P',S^{(1)} \mid T_i^{+-}(0) \mid PS^{(1)} \rangle & = &
- A_i(\Delta^2) i M_N \Delta^{(2)}  \nonumber \\
&~&~~ + {\overline C}_i(\Delta^2) ~ M_N~ g^{+-} ~ \Big ( (-) i
(\Delta^{(2)}  \Big )~. \label{pm}
\ee

From Eq. (\ref{pp}) we re-confirm that $A_i(\Delta^2)$ does not appear in the 
matrix element of $T^{++}$ in a transversely polarized state \cite{bhms}.

Methods I and  III yield the same results as Eqs.
(\ref{pp})-(\ref{pm})
in their dependence on $A_i$,
$B_i$ and ${\overline C}_i$. Note, however, that in these methods, $\Delta^-$
appears which we need to evaluate.   
Since $P$ and $P'$ are on mass shell,  $\Delta^-$  is related to 
$\Delta^+$ and $\Delta^\perp$ by
\be
\Delta^- = - \frac{\Delta^+}{({\overline P}^{+})^2 - (1/4) (\Delta^{+})^2 } 
~ \left (M^2 + \frac{1}{4} (\Delta^\perp)^2 \right) \Rightarrow - \Delta^+
\frac{M^2}{(P^{+})^2}.
\ee 
We ignore the $(\Delta^\perp)^2$ term since we are interested only in terms 
linear in $\Delta$. We also need to use the Eq. \ref{mf}.
 
 We summarize the results obtained so far,  as follows. \\ 
1) In $T^{++}$ matrix element, coefficient of $A_i$ form factor vanishes 
and hence it depends only on $B_i$ form factor. \\
2) $T^{+2}$ matrix element depends only on $\Delta^{+}$  explicitly and \\
3) $T^{+-}$ matrix element depends only on $A_i$ and ${\overline C}_i$ form
factors. 

\vskip .2in
{\bf Matrix elements of the Pauli-Lubanski operator $W_i^1$}
\vskip .2in
Substituting the results for individual matrix elements, in Eq.
(\ref{wspin})
we get
\be
\frac{\langle P S^{(1)} \mid W_i^1 \mid P S^{(1)} \rangle}
{\langle P S^{(1)} \mid P S^{(1)} \rangle} &= &
\frac{1}{2 P^{+}} \Bigg [ \frac{P^{+}}{2} \Big (2 A_i(0) + B_i(0) + 
2 {\overline C}_i(0) \Big )~ M_N 
+ \frac{P^{-}}{2} \frac{(P^{+})^2}{M_N} B_i(0) \Bigg ] \nonumber \\
& = & \frac{1}{2} M_N \Big ( A_i(0) + B_i(0) + {\overline C}_i(0) 
\Big )~.  
\ee

Thus the  matrix elements of $T^{+2}_i$ and $T^{+-}_i$ make comparable 
contributions
to the matrix element of $W^1_i$ in a transversely polarized state. The
matrix element of $T^{++}$ does not contribute to the matrix element of
total $W^1$.    
\vskip .2in
{\bf Comment on frame dependence of $W_i^1$ Matrix elements}
\vskip .2in
From the definitions of intrinsic spin operators, it is clear that 
in $P^\perp=0$ frame irrespective of the polarization $S$,
\be
M_N  \langle P S \mid {\cal J}_i^1 \mid P S \rangle &=& \langle P S \mid 
W_i^1 \mid P S \rangle \label{wt}\\
\langle P S \mid {\cal J}_i^3 \mid P S \rangle &=& \langle P S \mid J_i^3 
\mid P S \rangle \label{Jt}.
\ee
Note that in the appendix D of Ref. \cite{hmr2}, the calculation of the 
matrix element
of the intrinsic transverse spin operator in a transversely polarized
dressed quark state in an arbitrary reference frame is presented and the
frame independence is explicitly demonstrated.  
To calculate the RHS of the above equations in the $P^\perp=0$ only 
and simply claim that the results are frame independent, as the authors of 
Refs. \cite{jxyplb} and \cite{jxy} do, is naive and without
any basis as we demonstrate in the following.  
Extending the calculation presented in the last section for a frame with 
non-zero $P^\perp$ (i. e., not putting $P^\perp =0$ from the very beginning) 
one could easily 
show that even though LHS of the above equations are frame independent, while 
RHS are  not necessarily frame independent.  

Explicit calculation shows that,
for a transversely polarized nucleon, 
\be
\frac{\langle P S^{(1)} \mid {\cal J}_i^1 \mid P S^{(1)} \rangle}
{\langle P S^{(1)} \mid P S^{(1)}\rangle} &=& \frac{1}{2}
\big (A_i(0)+B_i(0)+{\overline C}_i(0)\big )
\ee
and 
for a longitudinally polarized nucleon
\be
\frac{\langle P S \mid {\cal J}_i^3 \mid P S \rangle}{\langle P S 
\mid P S \rangle} &=& \frac{1}{2}\big (A_i(0)+B_i(0)\big ).
\ee
which are frame independent.
We also get
\be
\frac{\langle P S \mid {\cal J}_i^1 \mid P S \rangle}{\langle P S 
\mid P S \rangle} &=& 0, \nonumber \\
\frac{\langle P S^{(1)} \mid {\cal J}_i^3 \mid P S^{(1)} 
\rangle}{\langle P S^{(1)} \mid P S^{(1)} \rangle}&=& 0~.\label{29}
\ee
 Results in Eq (\ref{29}) are frame independent and correctly represent the fact that
the
expectation value of the helicity in a transversely polarized nucleon must 
be zero
and the expectation value of intrinsic transverse spin 
in a longitudinally polarized 
nucleon must be zero. On the other hand, RHS of the corresponding equations 
 as obtained from 
Eq. (\ref{wt}) and Eq. (\ref{Jt}) 
are frame dependent and do not reflect the correct results:
\be
\frac{\langle P S \mid W_i^1 \mid P S \rangle}{\langle P S \mid P S \rangle} 
&=&  \frac{1}{2}\big (A_i(0)+B_i(0)\big ){\overline P}^1, \nonumber \\
\frac{\langle P S^{(1)}  \mid J_i^3 \mid P S^{(1)} \rangle }
{\langle P S^{(1)} 
\mid P S^{(1)} \rangle} &=& - B_i\frac{{\overline P}^1}{2M_N}~. 
\ee
We get vanishing RHS only in the frame $P^\perp =0$. 

\vskip .2in
{\bf Comparison of two methods}
\vskip .2in

In Refs. \cite{hmr1} and \cite{hmr2} we have presented a transverse spin sum
rule for the nucleon in QCD using light front dynamics and intrinsic (boost
invariant) transverse spin operators. The analysis relies
on the explicit structure of Poincare generators and hence depends on the
details of QCD dynamics since transverse spin operators in the light front
theory are interaction dependent. We were able to separate the operator into 
terms with and without explicit coordinate dependence. The latter could be
further separated into quark (${\cal J}^i_{II}$) and gluon parts 
(${\cal J}^i_{III}$). We have demonstrated \cite{hmr1,hmr2} that 
the nucleon matrix element of ${\cal J}^i_{II}$ is directly related to the 
integral of the well-known transverse polarized structure function $g_T$ and
the nucleon matrix element of ${\cal J}^i_{III}$ is directly related to the 
integral of the  gluon distribution function that appears in transverse
polarized hard scattering \cite{ji}. In the case of helicity, we have
demonstrated similar connections \cite{hk}, namely nucleon matrix element of
(${\cal J}^3_{q(i)}$) is directly related to the polarized structure
function $g_1$ and the nucleon matrix element of
(${\cal J}^3_{g(i)}$) is directly related to the gluon distribution
relevant to nucleon helicity \cite{jaffe}. Thus the physical content of our
sum rules are very transparent.

On the other hand, the sum rule discussed in Refs. \cite{jxy} and
\cite{jxyplb} contain form factors that parameterize the off-forward matrix
elements of the energy momentum tensor. The details of the dynamics remain 
hidden in this formalism. The separation into orbital and intrinsic spin
parts is not visible and relation of the sum rules to the quark and gluon 
helicity and transverse spin distribution functions that appear in various 
deep inelastic processes remain obscure.

\vskip .2in
{\bf Conclusions}
\vskip .2in

We have found  that (i) both the form factors $A_i$ and ${\overline C}_i$    
contribute
to the matrix element of $T_i^{+-}$ in a transversely polarized state, (ii)
there is no relative suppression factor between these two contributions and
(iii) 
the contribution to $W_i^\perp$  from $T_i^{++}$ contains only the form
factor 
$B_i$ and not the form factor $A_i$. 
First two observations differ from that in Ref. \cite{jxy} and eventually
invalidates their argument regarding the consequence of Lorentz invariance,
while the last finding is
already a  well established result \cite{bhms}. We have also compared and
contrasted the sum rules advocated by us with the sum rules following the 
method of Ref. \cite{jm}.  
\vskip .1in
{\bf Appendix}
\vskip .1in
All our calculations are performed in light front field theory. 
We will follow the conventions of Ref. \cite{pedestrian}.
Let us take the state to be polarized in the +ve x direction. 
Explicitly,  it is given by 
\be
\mid P, S^{(1)} \rangle = \frac{1}{\sqrt{2}} \Big ( \mid P, {\rm up} \rangle + 
\mid P, {\rm down} \rangle     \Big )
\ee
 where $\mid P, {\rm up} \rangle$ and 
$\mid P, {\rm down} \rangle$ are helicity eigenstates.  
Then we need to evaluate the matrix elements for up down, down up, up up and
down down helicity states. 

To simplify the calculations further, we use
\be
{\overline U}(P,S^{(1)}) \sigma^{\mu \nu} U((P,S^{(1)}) = 2 ~\epsilon^{\mu \nu \alpha
\beta} ~ \frac{P_\beta S_\alpha}{M_N}. \label{mf}
\ee  
Note that in our convention, $\epsilon^{+ - 1 2 }  = -2 $ and hence
Eq. (\ref{mf}) differs from the equation (5.36) of Ref. \cite{lbt} 
by a factor of 2. The components of the polarization vector
$S^\mu$ are explicitly $S^+=0$, $S^1 = M_N$, $S^{2}=0$ and $S^{-} = 2 M_N
\frac{P^1}{P^+}$.  
We present explicit calculations in Method II in which we need to calculate 
the five matrix 
elements, namely,
${\overline U}(P',S')
U(P,S)$ and ${\overline U}(P',S') \gamma^\mu  U(P,S)$. 

 Explicit evaluation of these matrix elements gives the following 
(in the frame with non-zero ${\overline P}^\perp$):
\vskip .1in
$\mathbf{ S^{\prime}=up, S=down}$ 
\be
{\overline U}(P',S') U(P,S) &  = & \frac{1}{\sqrt{P^+ {P'}^+}}\Big [ {\overline P}^+
\Big (\Delta^{(1)} - i \Delta^{(2)}  \Big )~ 
- \Delta^+
\Big ({\overline P^{(1)}} - i {\overline P^{(2)}}  \Big )~  \Big ], \\
{\overline U}(P',S') \gamma^1 U(P,S)  &=&  \frac{M_N}{\sqrt{P^+ {P'}^+}}
~\Delta^+~,  \\
{\overline U}(P',S') \gamma^2 U(P,S) & = & -i \frac{M_N}{\sqrt{P^+ {P'}^+}}
~\Delta^+~ \\
{\overline U}(P',S') \gamma^+ U(P,S)  &=&  0~,  \\
{\overline U}(P',S') \gamma^- U(P,S)  &=&  2 \frac{M_N}{\sqrt{P^+ {P'}^+}}~
\Big (\Delta^{(1)} - i \Delta^{(2)}  \Big )~.
\ee
\vskip .01in
$\mathbf{ S^{\prime}=down, S=up}$ 
\be
{\overline U}(P',S') U(P,S) & = & \frac{1}{\sqrt{P^+ {P'}^+}}\Big [ - {\overline P}^+
\Big ( \Delta^{(1)} + i \Delta^{(2)}  \Big )~
+\Delta^+
\Big ({\overline P^{(1)}} + i {\overline P^{(2)}}  \Big )~  \Big ],  \\
{\overline U}(P',S') \gamma^1 U(P,S) & = & - \frac{M_N}{\sqrt{P^+ {P'}^+}}
~ \Delta^+~,  \\
{\overline U}(P',S') \gamma^2 U(P,S) & = & -i \frac{M_N}{\sqrt{P^+ {P'}^+}}
~\Delta^+~,  \\
{\overline U}(P',S') \gamma^+ U(P,S) & = & 0~,  \\
{\overline U}(P',S') \gamma^- U(P,S) & = & - 2 \frac{M_N}{\sqrt{P^+ {P'}^+}}~
\Big ( \Delta^{(1)} + i (\Delta^{(2)}  \Big )~.
\ee
$\mathbf{ S^{\prime}=up, S=up}$ 
\be
{\overline U}(P',S') U(P,S) & = & \frac{1}{\sqrt{P^+ {P'}^+}} {\overline P}^+
\Big ( 2M_N  \Big )~,  \\
{\overline U}(P',S') \gamma^1 U(P,S) & = & 
\frac{1}{\sqrt{P^+ {P'}^+}}
~\Big [ 
{\overline P}^+   \Big ( 2{\overline P}^{(1)}- i\Delta^{(2)}  \Big )-\frac{\Delta^+}{2} \Big ( \Delta^{(1)} - 2i {\overline P}^{(2)}\Big ) 
\Big ],  \\
{\overline U}(P',S') \gamma^2 U(P,S) & = &  
\frac{1}{\sqrt{P^+ {P'}^+}}
~\Big [ 
{\overline P}^+\Big ( i\Delta^{(1)}+2{\overline P}^{(2)}  \Big )- 
\frac{\Delta^+}{2} \Big (2i{\overline P}^{(1)}+\Delta^{(2)}\Big ) 
\Big ],  \\
{\overline U}(P',S') \gamma^+ U(P,S) & = & 2 \sqrt{P^+ {P'}^+},  \\
{\overline U}(P',S') \gamma^- U(P,S) & = &  \frac{2}{\sqrt{P^+ {P'}^+}}~
\Big [  M_N^2 +({\overline P}^\perp)^2- \frac{1}{4} (\Delta^\perp)^2 \nonumber\\
&\quad &~~~~~~~~~~~~~~~~~~~~~~~~~~~~~~~~~~~~~~
+ i\Big ( {\overline P}^{(2)}\Delta^{(1)}-{\overline P}^{(1)}\Delta^{(2)}\Big )\Big ]~.
\ee
$\mathbf{ S^{\prime}=down, S=down}$ 
\be
{\overline U}(P',S') U(P,S) & = & \frac{1}{\sqrt{P^+ {P'}^+}} {\overline
P}^+
\Big ( 2M_N  \Big )~,  \\
{\overline U}(P',S') \gamma^1 U(P,S) & = & 
\frac{1}{\sqrt{P^+ {P'}^+}}
~\Big [{\overline P}^+ \Big ( 2{\overline P}^{(1)}
+i\Delta^{(2)}  \Big )
-\frac{\Delta^+}{2}\Big ( \Delta^{(1)} + 2i {\overline P}^{(2)}\Big )  
\Big ],  \\
{\overline U}(P',S') \gamma^2 U(P,S) & = &  
\frac{1}{\sqrt{P^+ {P'}^+}}
~\Big [{\overline P}^+\Big (- i\Delta^{(1)}+ 2{\overline P}^{(2)}  \Big )
\nonumber \\
&\quad &~~~~~~~~~~~~~~~~~~~~~~~~~~~~~~~~~~~~~- \frac{\Delta^+}{2} \Big (-2i{\overline P}^{(1)}+\Delta^{(2)}\Big )  \Big ],  \\
{\overline U}(P',S') \gamma^+ U(P,S) & = & 2 \sqrt{P^+ {P'}^+},  \\
{\overline U}(P',S') \gamma^- U(P,S) & = &  \frac{2}{\sqrt{P^+ {P'}^+}}~
\Big [  M_N^2 +({\overline P}^\perp)^2- \frac{1}{4} (\Delta^\perp)^2\nonumber\\
&\quad &~~~~~~~~~~~~~~~~~~~~~~~~~~~~~~~~~~~~~~ 
+ i\Big ( {\overline P}^{(1)}\Delta^{(2)} -{\overline P}^{(2)}\Delta^{(1)}\Big )\Big ]~.
\ee

\end{document}